\documentclass[letter]{aa}  
\usepackage{graphicx}
\usepackage{txfonts}
\usepackage{natbib}
\bibpunct{(}{)}{;}{a}{}{,}
\begin{document}
   \title{On the nature of X-Ray Flashes in the SWIFT era}

\author{B. Gendre\inst{1}$^{, }$\inst{2},
          A. Galli\inst{1}$^{, }$\inst{3}
	  \and L. Piro\inst{1}
          }

   \offprints{B. Gendre}

   \institute{IASF-Roma/INAF, via fosso del cavaliere 100, 00133 Roma, Italy\\
              \email{gendre@iasf-roma.inaf.it, galli@iasf-roma.inaf.it, piro@iasf-roma.inaf.it}
	\and
	Università degli Studi di Milano - Bicocca - Piazza dell'Ateneo Nuovo, 1 - 20126, Milano
         \and
             INFN-Trieste, Padriciano 99, 34012 Trieste, Italy
             }

   \date{Received ---; accepted ---}

 
  \abstract
   {}
   {X-Ray Flashes (XRFs) are soft gamma-ray bursts whose nature is not clear. Their soft spectrum can be due to cosmological effects (high redshift), an off-axis view of the jet or can be intrinsic to the source. We use SWIFT observations to investigate different scenarios proposed to explain their origin.}
   {We have made a systematic analysis of the afterglows of XRFs with known redshift observed by SWIFT. We derive their redshift and luminosity distributions, and compare their properties with a sample of normal GRBs observed by the same instrument.}
   {The high distance hypothesis is ruled out by the redshift distribution of our sample of XRFs, indicating that, at least for our sample, the off-axis and sub-energetic hypotheses are preferred. Of course, this does not exclude that some XRFs without known redshift could be at high distance. However we find that taking into account the sensitivity of the BAT instrument, XRFs cannot be detected by SWIFT beyond $\approx$ 3. The luminosity distribution of XRF afterglows is similar to the GRB one. This would rule out most off-axis models, but for the homogeneous jet model. However this model predicts a GRB rate uncomfortably near the observed rate of supernovae. This implies that XRFs, at least those of our sample, are intrinsically soft.}
   {}


   \maketitle
%

\section{Introduction}

A new class of ``soft Gamma-Ray Burst (GRB)'' making up about one third of the total population of GRBs was discovered by BeppoSAX \citep{heise}. An observational classification based on the hardness ratio of the prompt emission divides GRBs in : the classic hard GRBs, the soft X-Ray Flashes (XRFs), and an intermediate class of X-Ray Rich (XRR) events \citep{lamb03, barr}. In term of their spectral behaviors, XRFs showed a shape similar to the GRB one, with the only difference that the the energy at which the $\nu F_\nu$ spectrum peaks \citep[the peak energy, $E_p$, see][]{ban92} has a lower value \citep{sak05, dal06}. They fulfill the so-called Amati relation \citep{lamb05}.

Three scenarios have been proposed to explain the origin of XRFs. The high redshift scenario \citep{heise} appeared to be the most straightforward explanation for these events. In fact, a normal GRB placed at a redshift of 7-8 would be observed as an XRF due to the cosmological effects \citep{dal06}. Thus, the softness of XRFs could be only an observational bias due to distance.

The off-axis scenario is based on the assumption that we observe normal GRBs in (or very nearby to) the axis of the jetted fireball that produces the burst, while XRFs are observed off-axis. Several models, assuming different jet structures, have been proposed to account for the soft spectrum of XRFs \citep{yamaz02, yamaz03, yamazaki04, lamb05, zhang03, eichler, toma}.

Finally, the soft spectrum of XRFs could be due to an intrinsic property of the source (e.g. a sub-energetic or an inefficient fireball), that would radiate most of its energy in the X-ray band rather than in the gamma-ray one \citep{dermer, moc, ram}.

In a previous work using the BeppoSAX observations, \citet{dal06} have tested the distant event and off-axis observation hypotheses. Their main conclusion was that the dataset they had in hand gave little support to the high distance scenario. In fact, the measured redshifts of a few XRFs  ruled out the hypothesis that all of them are distant events, thus calling for an alternative explanation for nearby events. Under the assumption that GRBs and XRFs in their sample were at the same average distance, \citet{dal06} showed that a jet observed off-axis could marginally explain the data. However, they recognized that their findings need to be verified with an adequate sample of events with known redshift and therefore with a known luminosity distribution. 

The purpose of this letter is to further investigate the XRF nature, taking advantage of the larger sample of event with known distance available in the SWIFT era. We present the sample and the data analysis in section \ref{sec_ana}. While not all XRFs are distant events, some could be. We investigate the possible existence of distant events in section \ref{sec_red}. Using the luminosity distribution of XRFs, we discuss off-axis models in section \ref{sec_axis}. We finally summarize our findings and conclude in section \ref{sec_conclu}. In the following, we use a flat universe model ($H_0 = 73$, $\Omega_m = 0.23$).

\section{Data reduction, analysis and results}
\label{sec_ana}

 A burst is classified as an XRF when the Softness Ratio (SR) between the fluences in the 2.0-30.0 keV band to the 30.0-400.0 keV band is greater than 1 \citep{lamb03}. A burst whose SR is between 1 and 0.32 is classified as XRR, and included in the XRF sample by \citet{dal06}. Because the SWIFT \citep{ghe05} BAT instrument \citep{bar05}, which provide the trigger condition, has a narrower energy band (15-150 keV), we cannot access to SR directly. This narrow range also imply that the characteristic energy of the Band function can lie outside the detection band in several cases. In fact, most of the SWIFT prompt emission spectra are well fit by a simple power law \citep{zha06}, as one can expect in such a case. To build our XRF sample, we thus need to translate the condition on SR into a condition on the observed power law index. \citet{dal06} found that the mean values for their XRF sample were $E_p = 36$ keV, $\alpha = 1.2$, $\beta = 1.7$. We conservatively selected in our sample only those events satisfying the conditions $E_p < 15$ keV and $\beta > 2$, that give the condition SR $>$ 1.

We constructed our sample using events observed by SWIFT until the 12th of May 2006 (included). We retrieved the best fit power law indexes of the prompt spectra from the SWIFT official web page\footnote{see http://swift.gsfc.nasa.gov/docs/swift/archive/grb\_table/}. Twenty-one events were classified as XRFs using our criterion. We made a systematic analysis of the XRT observations of those events (details are given in Galli et al., in preparation). In this letter, we restrict our sample to bursts with known redshift (9 XRFs, 43 \% of the total sample). These events are compared to SWIFT GRBs with known redshift for which a published XRT light curve is available. Most of these bursts came from the work of \citet{obr05}. Tables \ref{table_xrf} and \ref{table_grb} list our samples of XRFs and GRBs respectively, together with the 2.0-10.0 keV fluxes (expressed 40 ksec after the burst, observer frame) and luminosities (expressed 20 ksec after the burst, rest frame). We compute this latter value using the flux observed at $20 \times (1+z)$ ksec, applying the k-correction using the spectral indexes listed in Galli et al. (in preparation).

\begin{table}
\caption[]{SWIFT XRF sample. We report the redshift, the prompt spectral parameter $\beta$, the 2.0-10.0 keV flux 40 ksec after the burst (observer frame) and the 2.0-10.0 keV luminosity 20 ksec after the burst (rest frame).}
         \label{table_xrf}
\centering                          
\begin{tabular}{ccccc}        
\hline\hline                 
XRF & Redshift & $\beta$ & X-ray flux                & X-ray \\  
name&          &         & $10^{-13}$ erg s$^{-1}$ cm$^{-2}$    &  luminosity \\ 
    &          &         & (40 ksec,                 & $10^{44}$ erg s$^{-1}$            \\
    &          &         & observer frame)           &  (20 ksec \\
    &          &         &                           & rest frame) \\
\hline
\object{060512} &0.4428 &  2.49   &    0.9                    & 0.1\\
\object{060218} &0.033  &  2.5    &    1.9                    & 0.00071\\
\object{060108} &2.03   &  2.01   &    3.2                    & 12.3\\
\object{051016B}&0.9364 &  2.38   &    6.8                    & 4.2\\
\object{050824} &0.83   &  2.78   &    5.2                    & 2.1\\
\object{050416A}&0.6535 &  3.20   &    6.0                    & 1.6\\
\object{050406} &2.44   &  2.44   &   0.57                    & 3.6\\
\object{050319} &3.24   &  2.38   &   20.0                    & 201\\
\object{050315} &1.949  &  2.11   &   29.0                    & 107\\
\hline
\end{tabular}
\end{table}

\begin{table}
\caption[]{SWIFT GRB sample. We report the redshift, the prompt spectral parameter $\beta$, the 2.0-10.0 keV flux 40 ksec after the burst (observer frame) and the 2.0-10.0 keV luminosity 20 ksec after the burst (rest frame). \object{GRB 050408} was detected by HETE-2; it has $E_p = 18$ keV, thus classifying this event as GRB using our criterion. It is however more likely an X-Ray Rich event rather than a normal GRB.}
         \label{table_grb}
\centering                          
\begin{tabular}{ccccc}        
\hline\hline                 
GRB & Redshift &$\beta$ & X-ray flux            & X-ray \\  
name&          &        &  $10^{-13}$ erg s$^{-1}$ cm$^{-2}$ & luminosity \\  
    &          &        &  (40 ksec,             & $10^{44}$ erg s$^{-1}$            \\
    &          &        &  observer frame)       &  (20 ksec \\
    &          &        &                        & rest frame) \\
\hline  
\object{050525A}& 0.606 &  1.06  &  8.51                  &  2.1\\
\object{050318} & 1.44  &  1.31  &  5                     &  2.5\\
\object{050223} & 0.5915&  1.84  &  0.691                 &  0.2\\
\object{050505} & 4.27  &  1.40  &  15.1                  &  111\\
\object{050126} & 1.29  &  1.33  &  0.375                 &  0.5\\
\object{050401} & 2.9   &  1.41  &  20.6                  &  119\\
\object{050408} & 1.236 & (HETE) &  8.88                  &  14.4\\
\object{051111} & 1.549 &  1.32  &  2.52                  &  7.2\\
\object{050730} & 3.97  &  1.52  &  38                    &  77.2\\
\object{060124} & 2.3   &  1.89  &  60.3                  &  307\\
\object{050603} & 2.281 &  1.17  &  12.2                  &  38.8\\
\object{050802} & 1.71  &  1.55  &  5.98                  &  12.4\\
\object{050922C}& 2.198 &  1.35  &  2.08                  &  14.7\\
\object{050820A}& 2.612 &  1.21  &  41.25                 &  255\\
\object{050908} & 3.35  &  1.86  &  0.2071                &  193\\
\object{050803} & 0.422 &  1.40  &  12.85                 &  1.1\\
\hline 
\end{tabular}
\end{table}

\begin{figure*}
\centering
\includegraphics[width=8.5cm]{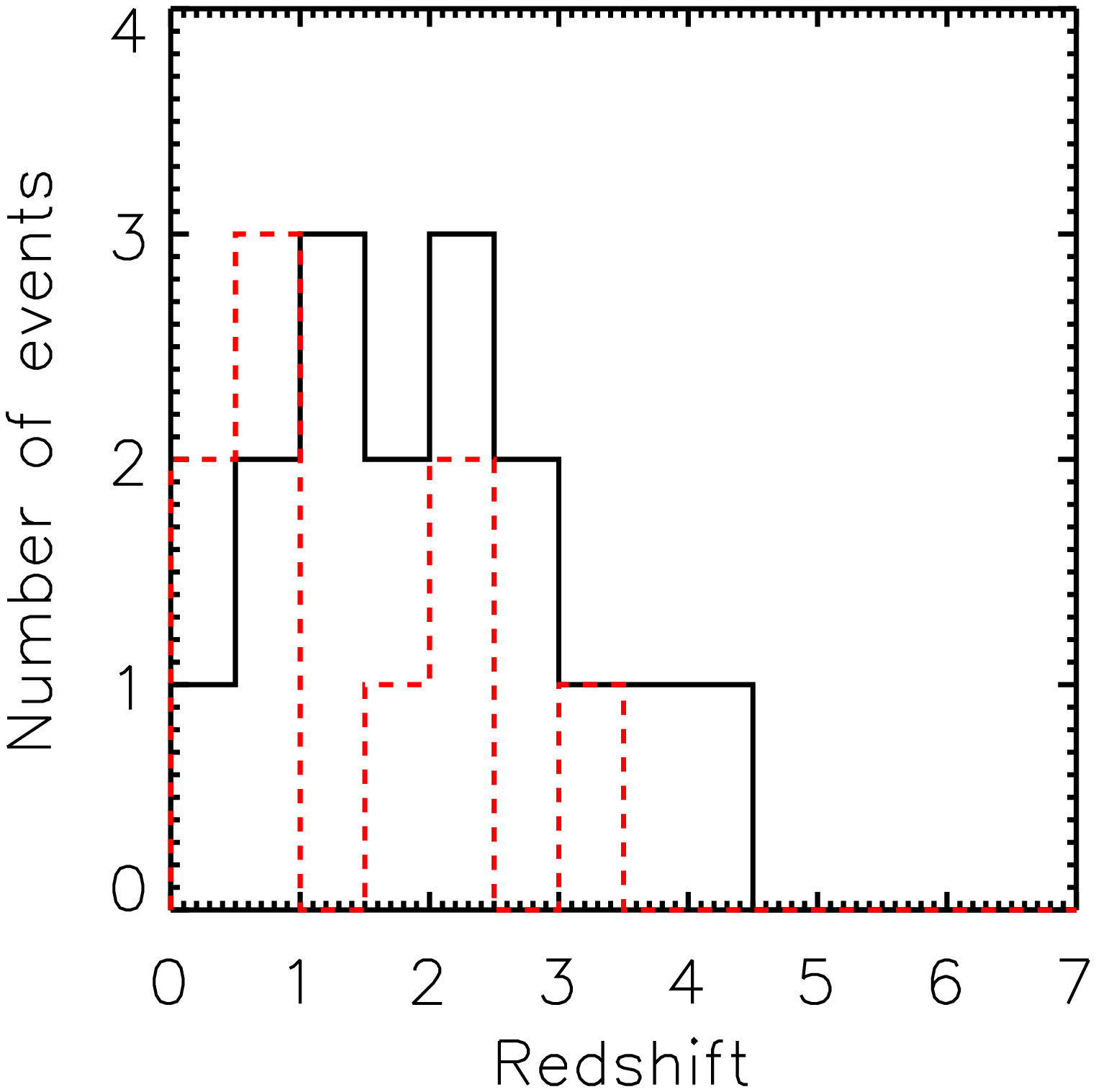}
\includegraphics[width=8.5cm]{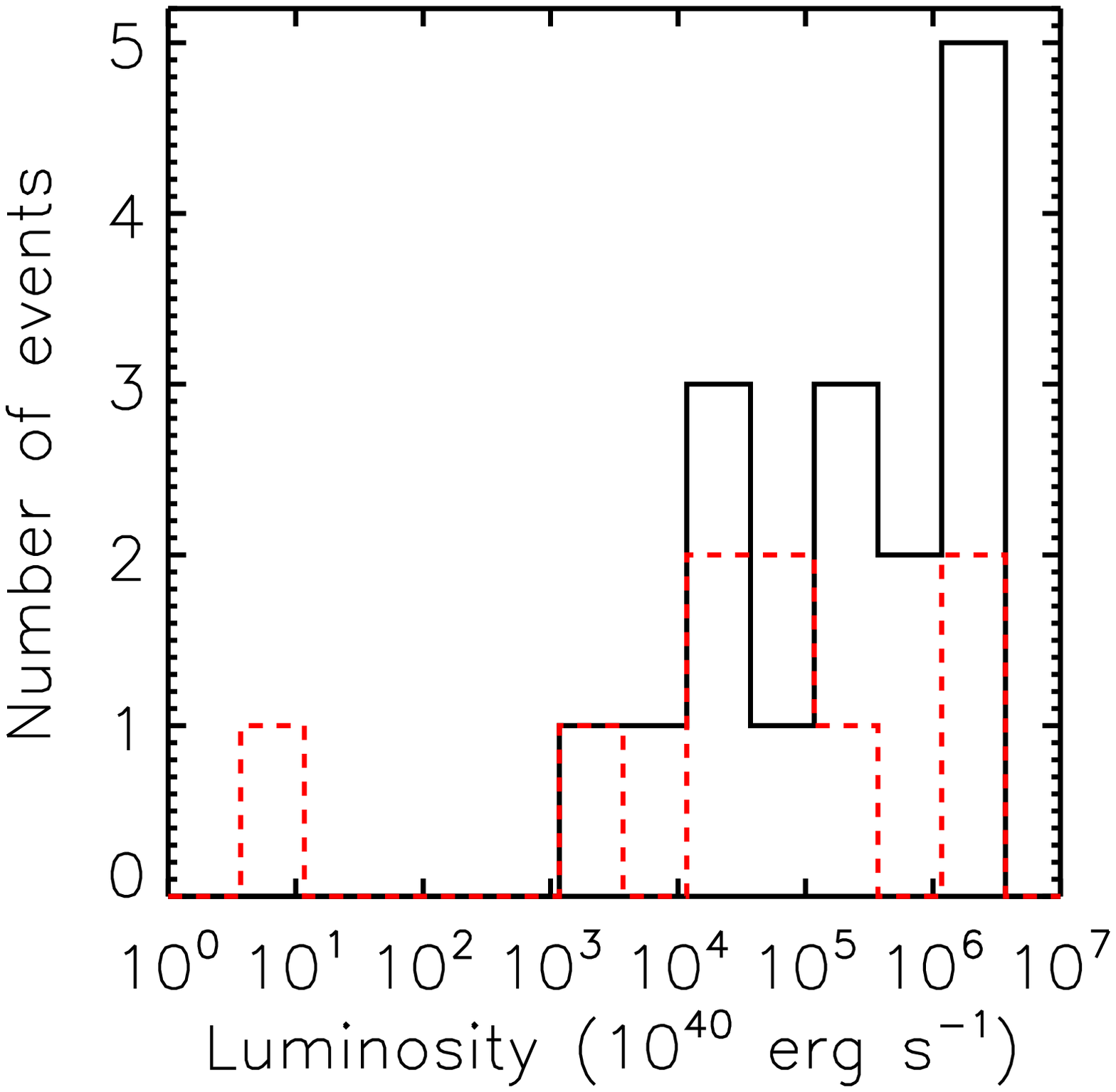}
\caption{The redshift (left) and luminosity (right) distributions of XRFs (dashed red line) and GRBs (solid black line). Luminosities are given at 20 ks in the burst frame.}
   \label{fig_flux}
   \label{fig_redshift}
\end{figure*}

The mean redshifts of SWIFT XRFs and GRBs are $\langle z_{XRF}\rangle = 1.40$, $\sigma = 1.06$ and $\langle z_{GRB}\rangle = 2.05$, $\sigma = 1.16$ respectively. The logarithmic mean fluxes of our two samples are $\langle Log F_{XRF, 13}\rangle = 0.6 \pm 0.2$ and $\langle Log F_{GRB, 13}\rangle = 0.8 \pm 0.2$. Correcting for distance effects, we obtain a logarithmic mean luminosity of $\langle Log L_{XRF, 44}\rangle = 0.34 \pm 0.55$ and $\langle Log L_{GRB, 44}\rangle = 1.18 \pm 0.25$. Note that XRF 060218 has a very low luminosity (see Fig. \ref{fig_flux} and Table \ref{table_xrf}). Excluding this event we obtain a mean luminosity $\langle Log L_{XRF, 44}\rangle = 0.77 \pm 0.37$, lower by a factor of $ 2.58$ compared to the mean GRB luminosity.


\section{The high distance hypothesis}
\label{sec_red}

The softness of XRFs compared to GRBs can be accounted for by assuming that XRFs are at higher ($z>5$) distance than GRBs \citep{heise}. No SWIFT XRFs with known redshift is at high distance (see Fig. \ref{fig_redshift}), thus implying that this hypothesis cannot account for all XRFs. However, some of them could indeed be high redshift GRBs. A distant event would have no optical afterglow emission because of the Lyman break redshifted in the optical band \citep{fru99}. If a significant fraction of XRFs are high distance GRBs, one may roughly expect a greater fraction of {\it dark} events among XRFs compared to GRBs \citep{dep04}. However, in our complete sample of 21 XRFs, twelve (57.1\% of the sample) have an optical afterglow (to be compared to the 48.7\% of SWIFT GRBs that have an optical afterglow)\footnote{see http://www.mpe.mpg.de/\~jcg/grbgen.html}. This would argue against the fact that a significant fraction of our XRF sample are high redshift GRBs. 

However, one may wonder if this is due to any selection effect, since the detection threshold of the BAT instrument changes depending on the observed $E_p$ value of the burst. The BAT sensitivity threshold to an XRF with $E_p \sim 15$ keV is $\sim 3$ ph cm$^{-2}$ s$^{-1}$ in the 1.0-1000.0 keV band \citep{ban06}, while it is $\sim 1$ ph cm$^{-2}$ s$^{-1}$ in the same band for a GRB with $E_p \sim 150$ keV. We have computed the minimum peak luminosity needed for an XRF with an observed $E_p \sim 15$ keV in order to trigger the BAT instrument as a function of the redshift, and compared it to the measured peak luminosity of our sample. 

\begin{figure}
\centering
\includegraphics[width=8.5cm]{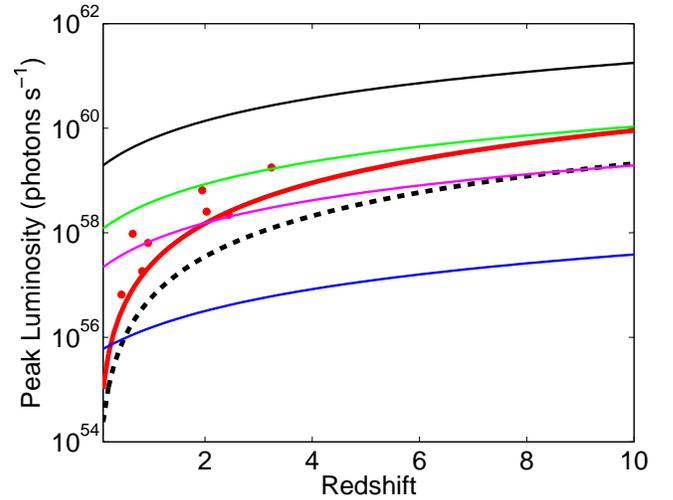}
   \caption{The selection effects. The solid red and dashed black lines represent the detection threshold of the SWIFT BAT instrument for an XRF with an observed $E_p = 15$ keV and a GRB with an observed $E_p = 150$ keV respectively. The red points are the peak luminosity of our sample. The other solid lines represent the peak luminosity expected from the Amati relation to produce a burst with an $E_p$ of 2 keV (blue line), 15 keV (purple line), 36 keV (green line) and 150 keV (black line) in the observer frame. See the electronic version for colors.}
      \label{fig_xrf2}
\end{figure}

Figure \ref{fig_xrf2} presents the result. One can clearly see that XRFs are too dim to be detected by SWIFT at high redshift. While several events are significantly above the calculated limit, an XRF that presents the mean characteristics of our sample could be detected only up to z $\sim 2$. Only a bright XRF could be observed up to z$\sim5-6$. In fact, using the Amati relation \citep{ama06} to compute the peak luminosity of a burst with an observed $E_p$ of 2 or 15 keV, and a duration of 20 seconds (rest frame), these events cannot be observed at redshift larger than $\sim 0.5$ and $\sim 2$ respectively. All of this imply that our sample is biased toward nearby and intermediate redshifts. This explains why we observe a mean redshift of XRFs lower than the GRB one. We stress that this effect is only due to the detection efficiency bias. Note however that SWIFT can indeed detect high distance GRBs, but these events have a very high intrinsic $\tilde{E}_p$ value which make them classified as GRBs rather than XRFs.

\section{The off-axis hypothesis}
\label{sec_axis}

\citet{dal06} have tested three different hypotheses for the jet geometry : the Universal Power law (UP), the Quasi-Universal Gaussian (QUG), and the Off-axis Homogeneous (OH) jets. These models involve a typical core size of the jet $\theta_c$, and predict a viewing off-axis angle ($\theta_v$) for a given intrinsic $\tilde{E}_p$ value. Using this value of $\theta_v$, one can then compute the afterglow emission light curve. Different $\tilde{E}_p$ values imply different viewing angles and thus different afterglow emission light curves for XRFs and GRBs. Large differences are expected at early times ($\lesssim 1$ day, rest frame) with the luminosity decreasing when $\theta_v$ increases. Hence, one can compute $R_X$, a GRB to XRF afterglow emission luminosity ratio at a given time \citep[see][for details]{dal06}. The $\tilde{E}_p$ values for the events of our samples are unknown. Using our criterion ($E_p = 15$ keV) and the mean XRF redshift, we obtain $\langle\tilde{E}_{p,~XRF}\rangle~ < 36$ keV. Six XRFs of our sample are listed in \citet{zha06}. Using their estimates, we have $\langle\tilde{E}_{p,~XRF}\rangle = 64.8$ keV. Note that this value is higher than the one implied by our assumption (see Galli et al. in preparation for a complete discussion); however this value is still lower than that of \citet{dal06}, $\langle\tilde{E}_{p,~XRF,~XRR}\rangle = 136$ keV. This latter value is higher because of the inclusion of XRRs in their sample. As for GRBs, using again the estimates of \citet{zha06} for our GRB sample, we derive $\langle\tilde{E}_{p,~GRB}\rangle \sim 398$ keV, in agreement with the BeppoSAX result of $\langle\tilde{E}_{p,~GRB}\rangle \sim 410$ keV \citep{dal06}.

Our sample of XRFs has a lower mean value of $\langle\tilde{E}_p\rangle$ compared to that of \citet{dal06}. Thus, the values of $R_X$ quoted in \citet{dal06} are lower limits for our sample. Assuming a core size of the jet large enough to account for the typical jet break time \citep[a few days,][]{dep06, gen06}, one should observe a ratio larger than $\sim 20$, $\sim 10$ and $\sim 1.2$ for the UP, QUG and OH models respectively, when expressed 20 ksec after the burst (rest frame). 

We observe $R_X = 2.6^{+2.6}_{-2.0}$. Because we now have access to the luminosity distribution rather than the flux one \citep[thus removing the uncertainties pointed out by][]{dal06}, we can clearly rule out large values of $R_X$. We thus confirm the claim of \citet{dal06} that an homogeneous jet with opening angle $\theta_j \simeq 6.1 \degr$ is the only one that can account for the observed $R_X$. Anyway a jet opening angle of $6 \degr$ fails to explain the distribution of prompt emission properties. In fact, population synthesis simulations of the bursts performed by \citet{lamb05} showed that, in order to reproduce the observed distribution of $E_{peak}$, $E_{iso}$ and fluence of a sample of GRBs plus XRFs detected by BeppoSAX and HETE-2, a mean opening angle $\theta_j \sim 0.5\degr$ is required. Furthermore, an OH jet with a such small mean opening angle implies a GRB/SN ratio of about 1. But, as shown by \citet{ber03} and \citet{sod04b}, no evidence for relativistic jet was found in SNe, thus constraining the GRB/SN fraction to be very low. Consequently, this jet model is also not favored by the global data.

\section{Discussion and conclusions}
\label{sec_conclu}

In this Letter, we have investigated the nature of XRFs, using a sample of events observed by SWIFT. As explained in Section \ref{sec_red}, the high distance hypothesis might hold for a very few (if any) of the soft events. In fact, SWIFT XRFs cannot be distant events due to a selection effect of the BAT instrument. The access to the luminosity distribution provided for the first time by SWIFT allowed us to strongly constrain the off-axis scenario. We found that for an suitable jet opening angle value an OH model is the only one that could account for the XRF X-ray afterglow luminosities. However, the same jet opening angle fails to account for prompt properties of XRFs versus GRBs.

One could put into question the core size of the jet, not well constrained yet. A lower value of the core size may reduce significantly the luminosity ratio expected in case of the UP model, thus making this model still plausible. However, the presence and occurrence time of temporal breaks in the afterglow emission light curve are key arguments to constrain the value of the core size of the jet, and will be discussed in Galli et al. (in preparation). A final possibility is that XRFs are produced by an inefficient fireball or a sub-energetic progenitor. For instance, in the context of internal shocks, XRFs may be produced by relativistic outflows with a low contrast in the Lorentz factor distribution, giving an efficiency of energy dissipation lower than in GRBs \citep{barr}. In an external shock scenario, a fireball with a low Lorentz factor can account for the low $E_p$ value observed \citep{dermer}. This can be produced by a fireball with a high baryon load or a sub-energetic progenitor. In that latter case, the model is however challenged to produce an under-luminous prompt emission but an XRF afterglow luminosity similar to the GRB one.

\begin{acknowledgements}
Thanks are due to E. M. Rossi for useful comments. We would also like to thank the anonymous referee for her/his very constructive report. BG acknowledge support from COFIN grant 2005025417.
\end{acknowledgements}

\end{document}